\documentclass[conference]{IEEEtran}
\usepackage{soul}
\usepackage{amsmath}
\usepackage{epsfig,epsf,psfrag,amssymb,amsfonts,latexsym,slashbox,graphicx,bm,cite,xcolor,url,array}
\usepackage{float}
\usepackage{caption}
\DeclareCaptionType{copyrightbox}
\usepackage{subcaption}
\usepackage{fixltx2e}
\usepackage{cases}
\usepackage{verbatim}
\usepackage[mathscr]{eucal}
\usepackage{algorithm, algorithmic}
\usepackage[fleqn,tbtags]{mathtools}
\usepackage{multirow}
\usepackage{enumitem}
\usepackage{paralist}
\usepackage[pass]{geometry}
\usepackage{color}

\DeclareMathOperator{\dep}{d}

\DeclareMathOperator{\Var}{Var}

\DeclareMathOperator{\Cov}{Cov}

 \def\0{{\bf 0}}

\def\qed{\hfill\hbox{${\vcenter{\vbox{
    \hrule height 0.4pt\hbox{\vrule width 0.4pt height 6pt
    \kern5pt\vrule width 0.4pt}\hrule height 0.4pt}}}$}}

\definecolor{myred}{rgb}{0.3,0.0,0.7}
\definecolor{dkg}{rgb}{0.1,0.7,0.2}
\definecolor{dkb}{rgb}{0.0,0.2,0.8}

\def\bfX{{\mathbf X}}

\def\thetabf{{\mbox{\boldmath$\theta$\unboldmath}}}

\def\Ec{{\cal E}}

\def\Tc{{\cal T}}

\def\Zc{{\cal Z}}

\newcommand{\bprf}{\begin{myproof}}
\newcommand{\eprf}{\end{myproof}}
\newcommand{\bp}{\begin{psfrags}}
\newcommand{\ep}{\end{psfrags}}
\newcommand{\bl}{\begin{lemma}}
\newcommand{\el}{\end{lemma}}
\newcommand{\bt}{\begin{theorem}}
\newcommand{\et}{\end{theorem}}
\newcommand{\bc}{\begin{center}}
\newcommand{\ec}{\end{center}}
\newcommand{\bi}{\begin{itemize}}
\newcommand{\ei}{\end{itemize}}
\newcommand{\ben}{\begin{enumerate}}
\newcommand{\een}{\end{enumerate}}
\newcommand{\bd}{\begin{definition}}
\newcommand{\ed}{\end{definition}}
\def\beq{\begin{equation}}
\def\eeq{\end{equation}\noindent}
\def\beqn{\begin{eqnarray}}
\def\eeqn{\end{eqnarray} \noindent}
\def\beqnn{  \begin{eqnarray*}}
\def\eeqnn{\end{eqnarray*}  \noindent}
\def\bcase{  \begin{numcases}}
\def\ecase{\end{numcases}   \noindent}
\def\bsbcase{  \begin{subnumcases}}
\def\esbcase{\end{subnumcases}   \noindent}

\newtheorem{theorem}{Theorem}

\newtheorem{lemma}{Lemma}

\newtheorem{definition}{Definition}

\newenvironment{myproof}{\noindent{\em Proof:} \hspace*{1em}}{
    \hspace*{\fill} $\Box$ }
\newenvironment{proof_of}[1]{\noindent {\em Proof of #1: }}{\hspace*{\fill} $\Box$ }

\newcommand{\matplottc}[1]{               
        \unitlength .45truein
        \begin{center}
        \includegraphics{#1.ps}
        \end{picture}
        \end{center}
}

\def\psfancypar#1#2{\begingroup\def\par{\endgraf\endgroup\lineskiplimit=0pt}
               \setbox2=\hbox{\large\sc #2}
               \newdimen\tmpht \tmpht \ht2 \advance\tmpht by \baselineskip
               \font\hhuge=Times-Bold at \tmpht
               \setbox1=\hbox{{\hhuge #1}}
               \count7=\tmpht \count8=\ht1
               \divide\count8 by 1000 \divide\count7 by \count8
               \tmpht=.001\tmpht\multiply\tmpht by \count7
               \font\hhuge=Times-Bold at \tmpht
               \setbox1=\hbox{{\hhuge #1}}
               \noindent
                \hangindent1.05\wd1
               \hangafter=-2 {\hskip-\hangindent
               \lower1\ht1\hbox{\raise1.0\ht2\copy1}%
                \kern-0\wd1}\copy2\lineskiplimit=-1000pt}

\def\Kout{\setbox1=\hbox{\Huge\bf K}\hbox to
1.05\wd1{\hspace{.05\wd1}
}}
\def\Sout{\setbox1=\hbox{\Huge\bf S}\hbox to 1.05\wd1{\hspace{.05\wd1}
}}

\allowdisplaybreaks[4]

\def\Ec{{\cal E}}

\newcommand{\clgroup}{\mathsf{CLGrouping}}
\newcommand{\estone}{\mathsf{LocalCLGrouping}}

\newcommand{\CondP}{\mathsf{CP}}
\newcommand{\CA}{\mathsf{CA}}

\newcommand{\EAp}{\mathsf{EP}}
\newcommand{\EAa}{\mathsf{EA}}

\newcommand{\RDA}{\mathsf{RDA}}

\newcommand{\RDM}{\mathsf{RDM}}

\begin{document}
%
\title{Are you going to the party: depends, who else is coming?\\
			\large [Learning hidden group dynamics via conditional latent tree models]}



%
\author{\IEEEauthorblockN{Forough Arabshahi\IEEEauthorrefmark{1},
Furong Huang\IEEEauthorrefmark{2},
Animashree Anandkumar\IEEEauthorrefmark{3}, 
Carter T. Butts\IEEEauthorrefmark{4} and
Sean M. Fitshugh\IEEEauthorrefmark{5}}
\IEEEauthorblockA{\IEEEauthorrefmark{1}farabsha, \IEEEauthorrefmark{2}furongh, \IEEEauthorrefmark{3}a.anandkumar, \IEEEauthorrefmark{4}buttsc, \IEEEauthorrefmark{5}sean.fitzhugh @uci.edu\\
University of California, Irvine}
}


\maketitle

\vspace{-2em}

\begin{abstract}
Scalable probabilistic modeling and prediction in high dimensional multivariate time-series is a challenging problem, particularly for systems with hidden sources of dependence and/or homogeneity. Examples of such problems include dynamic social networks with co-evolving nodes and edges and dynamic student learning in online courses. Here, we address these problems through the discovery of hierarchical latent groups. We introduce a family of Conditional Latent Tree Models (CLTM), in which tree-structured latent variables incorporate the unknown groups. The latent tree itself is conditioned on observed covariates such as seasonality, historical activity, and node attributes. We propose a statistically efficient framework for learning both the hierarchical tree structure and the parameters of the CLTM. We demonstrate competitive performance in multiple real world datasets from different domains. These include a dataset on students' attempts at answering questions in a psychology MOOC, Twitter users participating in an emergency management discussion and interacting with one another, and windsurfers interacting on a beach in Southern California. In addition, our modeling framework provides valuable and interpretable information about the hidden group structures and their effect on the evolution of the time series.
\end{abstract}


\section{Introduction}
In this work we address the problem of modeling and predicting high dimensional time-series with latent dependence and/or unobserved heterogeneity. Such time series arise in numerous important applications, including dynamic social networks with co-evolving nodes and edges, and dynamic student learning in MOOCs. Of particular interest in modeling such high dimensional series is the problem of predicting their evolution. Such predictions can in turn be used to provide useful feedback such as recommendations to network participants or students to improve their experience in the network and help them learn the course material (respectively). Modeling and tracking such high dimensional series jointly, however, is a greatly challenging task since each sequence can interact with others in unknown and complex ways. Before delving into the details of the prediction model, we thus first identify several factors that influence the dynamics of high-dimensional time series in a social context.

First and foremost, individual-level behavioral variables in a multivariate time series are strongly influenced by group dynamics. For example, the nodes in a social network tend to participate in communities, and the evolution of node behavior can be captured in part by the dynamics of those communities. In some cases, the resulting dependence is endogenous: for instance, a network attendee might wonder who else is going to attend a social event (e.g., a party) before deciding whether to attend him or herself. In other cases, group-level dependence may stem from unobserved heterogeneity: in the student learning scenario, for instance, students may be divided into groups of strong and weak learners whose learning curves evolve in drastically different ways. Hence, finding such underlying groupings and considering their dynamics for predicting the evolution of each individual sequence is of great importance. A second challenge for modeling in this context (as implied by the first example above), is that the dynamic behavior of each random variable affects the dynamics of other random variables, making the individual sequences dependent on one another. Treating each individual sequence independently ignores such interdependence and results in poor predictions.  A third challenge for modeling in this context is the need to account for the impact of relevant external factors (\emph{covariates}) that are predictive of dynamics. Seasonal or period effects are examples of covariates whose states can be predictive of the evolution of the series. E.g. in weekly social events, the day of week is a highly predictive factor of the attendance dynamics of the participants. Another example of relevant covariates in the context of student learning is the topic of each lesson or problem being studied, as each student has topic specific learning strengths and weaknesses. Last but not least, consecutive time points are highly correlated in typical time-series contexts. Therefore, knowing the previous state of the variables (and appropriately handling inertia) is vital for making good predictions.

Here, we introduce a parametric model class, namely the Conditional Latent Tree Models (CLTM), that takes into account the effect of all the above factors for predicting high-dimensional time-series. The effect of the covariates and previous time points is captured via Conditional Random Fields (CRF's). More specifically, conditioned on exogenous covariates and previous time points, the dependency structure among the variables is modeled via a latent tree whose hidden nodes represent the unobserved hidden groupings in the data. Therefore, CLTM represents the joint distribution of the observed and latent random variables which factorizes according to a Markov latent tree conditioned on the covariates and previous time points. This model is versatile in its ability to model group structure and provide the ability to carry out exact {\em inference} through the simple {\em belief propagation }(BP) algorithm~\cite{choi2011learning} that makes the model potentially scalable. We provide a statistically efficient algorithm for learning the structure of the latent tree, and estimate the parameters of the model using a Maximum Likelihood (ML) approach. Therefore, our goal in the sequel is to learn unobserved groups of similar behavior and incorporate them into prediction of the evolution of high dimensional time-series conditioned on some relevant covariates.

It is worth mentioning that like many common alternatives (e.g., hidden Markov models), the latent tree structural assumption used here is a reasonable approximation to the true dependence structure underlying the random variables arising from typical social settings. This approximation is obviously more realistic than a purely independent model, but also captures subtle hierarchical features that are missed or obscured by alternatives such as latent state models. It is also denser and more flexible than a simple tree over the observed variables due to the presence of the latent variables (since if we marginalize out the latent variables, the structure will not remain a tree anymore). Although we do not claim that latent tree structures are perfect representations of the myriad sources of dependence in large, complex social systems, we thus do regard the CTLM as an effective ``middle ground'' between simple independence and/or latent state models and difficult-to-scale models with unbounded dependence (e.g. full temporal ERGMs with endogenous attributes \cite{hanneke10}). As we show here, predictive results obtained by applying these models to several real-world data sets provide further evidence for their efficacy.


\subsection{Summary of Results}
\label{seq:ResSumm}
In this paper we introduce CLTMs and propose a framework for learning them efficiently. This framework has the potential to be applied to large scale data sets. We first estimate the latent group structure among the variables, and then learn the parameters of the CLTM, which describe quantitatively how the hidden variables affect the observed outcomes. We then employ CLTM's to efficiently track the evolution of time series. 



We apply our approach to three challenging real-world datasets involving students' performance in a psychology MOOC, Twitter users' activities and interactions, and windsurfers' participation in and interactions during activity on a southern California beach. In all these data sets our goal is to predict the dynamics of the users (either the students in the class or the network attendees) for which we need to extract relevant covariates that will be used in CLTM.

For the MOOC data, for example, In order to acquire predictive covariates for student prediction we first learn a conditional latent tree model over the knowledge components. Each question answered by any student incorporates a certain knowledge component. By learning the CLTM over the knowledge components, we can automatically find hierarchical groups of concepts that are learnt similarly by students. We demonstrate that the learned tree structure captures interpretable groupings. For instance, knowledge components related to different anxiety disorders are grouped together. Note that we only use these labels of knowledge components for validation, and not during the learning phase. We then incorporate these knowledge groupings as covariates in a CLTM used for tracking the learning of individual students over time. In this CLTM over the students, each observed node indicates the performance of one student on questions answered daily. Our model automatically learns groups of students who demonstrate similar evolution of learning behavior. Such information can be valuable to an instructor, since it gives him/her the ability to target different groups of students, and tune the instruction accordingly. Our approach is in contrast to earlier modeling frameworks for this MOOC dataset, which fit a different latent variable model for each student separately, treating the students independently, in order to model the learning progress~\cite{bier2014approach}.

We quantitatively compare the prediction of student performance under our method with a chain CRF model in Table \ref{tab:Res}, in which the chain is over time and the students are treated as independent time sequences similar to \cite{bier2014approach}.
We observe a significant improvement in predicting the student performance. Similarly, we also demonstrate a strong improvement on Twitter and beach data for predicting the conditional presence of vertices and edges over time. This is especially relevant, since these datasets are highly sparse with a small number of participants at any given time. Moreover, we observe that our method has higher improvement on the Twitter dataset compared to the beach data, since the beach dataset has covariates that are carefully collected by a team of sociologists. Thus, our method is highly effective in predicting multivariate time series across multiple domains, particularly where covariate information is present but limited.




\subsection{Related Work}
\label{seq:Related}
Previous works on multi-variate time series typically do not consider latent groups, e.g.~\cite{hamilton1994time}. This results in too many unknown parameters and results in the problem of overfitting and computational intractability in the high dimensional regime. The alternative is to first learn the groups through standard clustering techniques such as agglomerative clustering~\cite{krishnamurthy2012efficient}, and then use them as covariates for prediction. However, this two step process is not optimal for prediction. In contrast, our CLTM is a statistical model which simultaneously learns the groups and their effect on time evolution, leading to efficient performance.

Another interesting line of related work considers community models such as stochastic block models and mixed membership models~\cite{xing2010state,yang2014community,white2014mixed,anandkumar2013tensor,wallace2012extracting} for modeling the unknown vertex groups. However, these models only consider the edge data and do not incorporate node state information and exogenous factors. In our datasets, we also have node activity information (such as number of tweets by a user), and we exploit this information to learn about the unknown node groups. We then incorporate the group structure for learning the edge dynamics.  Further, the aforementioned works mostly  assume data samples to be independent and identically distributed (except for~\cite{wallace2012extracting} or \cite{foulds2011dynamic}), whereas we consider time varying data.


%

CLTM belongs to the class of Conditional random fields (CRF). 
Various CRFs have been considered before, e.g. CRFs on linear chains \cite{sha03}, trees \cite{quattoni2007hidden, bradley+guestrin:icml10crfs},  grids \cite{kumar03}, and so on. However, only a few works address the issue of structure learning of CRFs, e.g.~\cite{Torralba04contextualmodels,Schmidt08structurelearning,bradley+guestrin:icml10crfs}. Moreover, not many publications assume CRFs with latent variables, e.g~\cite{quattoni2007hidden} has latent variables, but with a fixed structure. Our work, on the other hand, does not make such strong assumptions. We learn the latent tree structure through efficient methods and also incorporate covariate effects, leading to highly effective models in practice.



\section{Model}
\label{sec: methodology}
Let us denote random variables with $y_i^{(t)} \in \mathbb{R}$ where $i = 1,2,\dots,n$ indicates the index of the random variable and $t = 1,2,\dots,T$ is the time index. We use the terms ``random variable'' and ``node'' interchangeably as the random variables can be represented as nodes in a dependency graph. An example of such a dependency graph is shown in Fig \ref{fig:toyStruct}. 

Each random variable's behavior is dependent upon other random variables' behavior, as well as a set of covariates. 
Examples of covariates include network users' group memberships, seasonal effects (e.g. the day of the week). There are three types of covariates we consider in this paper: the individual covariates which are node specific, e.g. membership of a specific node in an observed group (red nodes in Fig \ref{fig:toyStruct}); the shared covariates which indicate the dependency among the nodes; and the global covariates which simultaneously affect every node in the conditional latent tree, such as seasonality (the black node in Fig \ref{fig:toyStruct}). 
Let ${\bf x}_i^{(t)} \in \mathbb{R}^{(1 \times Kn)}$ indicate the set of node specific and global covariates, and ${\bf x}_{ij}^{(t)} \in \mathbb{R}^{(1 \times Ke)}$  indicate the shared covariates.
In this case, $Kn$ indicates the number of covariates of $y_i^{(t)}$ 
and $Ke$ indicates the number of shared covariates between nodes $y_i^{(t)}$ and $y_j^{(t)}$.

\subsection{Conditional Latent Tree Models (CLTM)}
\label{sec:CLTM}
Our goal is to perform structured prediction when there are temporal dynamics in the data. Consider an online social network such as Twitter as an example. Let the the network users be the random variables whose Tweeting activity is tracked over time. If two users are similar (say, both belong to a subgroup with similar demographic and social characteristics), it is more likely for them to have similar activity. Additionally, we claim that the users' attendance behavior depends on their previous activity, the behavior of other users and some relevant covariates. 
Note that throughout the paper, previous observations are contained in the prediction model as a subset of the covariates.
Therefore, we learn a latent tree dependence model over the users conditioned on the covariates and previous observations and we predict users' attendance dynamics according to the learned structure. As another example, we predict the performance of students in a course. Based on the students' performance on their exams throughout the semester we find groups of students who share similar learning behaviors using CLTMs. We will see that finding these similarities and hidden groupings can greatly help in predicting students' learning performance.

Let us first look at what the latent tree structure looks like and why we are assuming such a dependence structure. Consider the Twitter network in which $y_i$'s are the Twitting activity of network users, or an online education system in which $y_i$'s are the performance of the students on various course material such as problems and quizzes. The latent nodes in the tree are denoted by $h_j$ where $j=1,2,\dots,m$. They represent hidden groupings in the dependence structure of the random variables $y_i$. Let $z_k$ be the union of the observed nodes $y_i$ and latent nodes $h_j$ where $k = 1,2,\dots,n+m$. Let us denote the latent tree by $\Tc_{\dep} = (\Zc_{\dep},\Ec_{\dep})$ where $\Zc_{\dep}$ indicates the node set consisting of all the random variables and $\Ec_{\dep}$ denotes the edge set containing the edges of the latent tree. There are two main advantages in making latent tree structural assumptions. Firstly, a latent tree allows for more complex structures of dependence compared to a fully observed tree - specifically, it allows for latent groups of individuals whose behaviors jointly covary. Secondly, inference on it is tractable, and therefore, it will be scalable.

\begin{figure}[t]
	\centering{
	\psfrag{h1}[Bl]{\scriptsize  ${h_1}$}	
	\psfrag{h2}[Bl]{\scriptsize  ${h_2}$}
	\psfrag{x1}[Bc]{\scriptsize $x_1$}	
	\psfrag{x2}[Bc]{\scriptsize $x_2$}	
	\psfrag{x3}[Bc]{\scriptsize$x_3$}
	\psfrag{x4}[Bc]{\scriptsize $x_4$}	
	\psfrag{x5}[Bc]{\scriptsize $x_{h_1}$}
	\psfrag{x6}[Bc]{\scriptsize $x_{h_2}$}
	\psfrag{x7}[Bc]{\textcolor[rgb]{1,1,1}{\scriptsize $x_{g}$}}
	\psfrag{y1}[Bl]{\textcolor[rgb]{1,1,1}{\scriptsize ${y_1}$}}	
	\psfrag{y2}[Bl]{\textcolor[rgb]{1,1,1}{\scriptsize ${y_2}$}}	
	\psfrag{y3}[Bl]{\textcolor[rgb]{1,1,1}{\scriptsize  ${y_3}$}}	
	\psfrag{y4}[Bl]{\textcolor[rgb]{1,1,1}{\scriptsize  ${y_4}$}}
	\includegraphics[width = 0.25\textwidth]{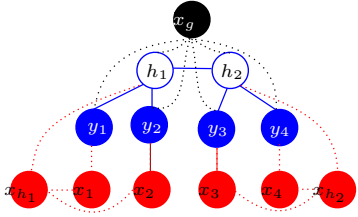}}
	\caption{Toy example of the underlying structure over the random variables. Blue nodes are observable variables, blank nodes are hidden variables, red nodes denote the individual covariates and the black node denotes the global covariate. As shown in the figure, the dependence structure of the random variables is a tree conditioned on covariates.}
	\label{fig:toyStruct}
	\vspace{-2em}
\end{figure}

Once the dependency structure is achieved, we should specify the generative distribution that the data is drawn from. The distribution of the random variables in CLTM belongs to the exponential family conditioned on observed covariates $X$. Covariates are observable external factors that affect the dynamics of the data. For example, in the Twitter network, relevant covariates are seasonality, regularity of network users, their popularity, and their previous activity. Fig \ref{fig:toyStruct} demonstrates the CLTM structure conditioned on the covariates $X$. As one can see in the figure, the joint structure of the observed and hidden variables are that of a tree conditioned on the covariates. Now let us give more details about the data distribution.

Exponential family distributions are a broad family of distributions including the normal, Gamma, Poisson and many other distributions \cite{gupta2001exponentiated}. Conditioned on covariates $X$, the distribution of $Z$ over tree $\Tc_{\dep}$ is given in Equation \eqref{eq:expDist}.
\begin{equation}\label{eq:expDist}
 \text{Pr}(Z \vert X, \thetabf)\hspace{-0.5em}=\hspace{-0.3em}\exp\hspace{-0.5em}{\left(\hspace{-0.3em} \sum\limits_{k \in \mathcal{Z}_{\dep}}\phi_k( X,\thetabf )z_k \hspace{-0.4em}+ \hspace{-0.7em}\sum_{kl \in \Ec_{\dep}} \phi_{kl}( X ,\thetabf )z_{k} z_{l} \hspace{-0.4em}- \hspace{-0.5em}A\hspace{-0.3em}\left(X, \thetabf\right) \hspace{-0.5em}\right)},
\end{equation}
where $A(X, \thetabf)$ is the term that normalizes the distribution, also known as the log partition function. $\phi_k( X,{\bf\theta} )$ and $\phi_{kl}( X ,{\bf\theta} )$ indicate the node and edge potentials of the exponential family distribution, respectively. Let's assume for the sake of simplicity that the potentials are linear functions of the covariates and previous observation as shown below.
\begin{equation}
\label{eq:nPot}
\phi_k( X,\thetabf) = c_0 + c_1 x_{1,k}+ ... + c_{K_n} x_{P,k},
\end{equation}
\begin{equation}
\label{eq:ePot}
\phi_{kl}( X,\thetabf) = e_0 + e_1 x_{1,kl} ... + e_{K_e} x_{P,kl},
\end{equation}
Learning the graphical model involves two steps: learning the dependence structure over the nodes, and estimating the probability distribution the data is generated from. In the following paragraphs we provide detailed description of these two steps.

The details of estimating the distribution is presented in Section \ref{sec:paramEst}. It should be noted that the random variables in this model can either be discrete or continuous and we will cover both variables in the structure learning and parameter estimation sections.
\subsection{Exploiting Inferred Hidden Groups}
\label{sec:EdgePred}
A problem of frequent current interest is that of modeling the dynamics of data on social interactions, as e.g. occur in online social networks. E.g., in the context of the Twitter network a social tie (an edge) can represent any form of communication between the users while they are active. We learn these dynamics by using the information from the learnt CLTM model for node activity. It is reasonable to claim that an edge cannot be formed unless both parties that form it are present in the network. In that case, if we correctly predict the active users in the network, our chances of predicting edge dynamics increase. Therefore, conditioned on the model learned for the users we predict users' interaction patterns by regressing on a set of relevant edge covariates and the inferred state of the hidden variables from the CLTM model.

Let $W$ represent the set of edges (social ties), and $w_{ij}$ denote the presence/absence of an edge between users $i$ and $j$. Intuitively, by regressing on the inferred hidden states, we can incorporate the latent group dynamics into prediction of edges. Assuming independence between the edges, the generative model of the edges given user activity, the inferred state of the hidden variables obtained from the CLTM model, and relevant edge covariates $X$ is:
\begin{equation}
	\label{eq:edge}
	\Pr(W\mid Z, X ) = \prod_{(i,j)} B\left(w_{ij}\left\lvert \textrm{logit}^{-1} \left(\xi(Z,X) \right)\right.\right), 	
\end{equation}where $\xi(Z,X)$ is a linear function of the the edge covariates $X$, consisting of past network information, current node states $Y$ and inferred states of the hidden variables of the CLTM as shown in Equation \eqref{eq:adjCov}. $B$ is the Bernoulli distribution and $\text{logit}^{-1}$ is the logistic function. Note that re-use of variable $X$ for the edge covariates is due to notational simplicity and in practice, the edge and node covariates are not the same. $X_{k,ij}^{(t)}$ in the equation below denotes the $k^{th}$ covariate for the edge formed between nodes $i$ and $j$, at time point $t$.
\begin{equation}
	\label{eq:adjCov}
	\xi(Z,X) =	d_0 + d_1 X_{1,ij}^{(t)}+ \dots + d_{K_{ec}} X_{K_{ec},ij}^{(t)} + {\bf d}_n Z.
\end{equation}

\section{Model Estimation}
\label{sec: estimation}
In order to be able to do prediction, we first estimate the underlying model given some observations. Model estimation can be divided into two general steps: structure learning and parameter estimation. For structure learning, the target structure is the structure of a latent tree that will be learned using a notion of information distances efficiently. For parameter estimation, we solve the exponential family distribution defined in Equation \eqref{eq:expDist} using maximum likelihood. EM is used in the parameter estimation step as latent nodes introduced in the tree by structure learning step are not observed. 

\subsection{Structure Learning}
\label{sec:structLearn}
A large number of scalable structure learning algorithms for latent tree modeling have been discovered by the phylogenetic community on learning latent tree models. Among the available approaches, we build upon \textsf{RG} and $\estone$ \cite{choi2011learning} with provable computational efficiency guarantees. These algorithms are based on a measure of statistical additive tree distance metric $d$ (a.k.a information distance) which is given below for discrete random variables:
\begin{align}
\label{eq:dist}
d_{ij}:= - \log \frac{\vert \det {\bf J}(y_i,y_j)\vert}{\sqrt{\det {\bf M}(y_i) \det {\bf M}(y_j)}},
\end{align}
where ${\bf J}(y_i,y_j)$ is the joint probability matrix between $y_i$ and $y_j$ and ${\bf M}(y_i)$ is the diagonal marginal probability matrix of node $y_i$. In practice, we employ empirical estimates of ${\bf J}(y_i,y_j)$, ${\bf M}(y_i)$ and ${\bf M}(y_j)$ based on sampled data. 
For continuous Gaussian random variables the information distances are the log of their correlation coefficient:
\begin{equation}
\label{eq:distCont}
 d_{ij} := -\log \frac{\Cov(y_i,y_j)}{\sqrt{\Var(x_i)\Var(x_j)}}.
\end{equation}
The distance measure given in Equations \eqref{eq:dist} and \eqref{eq:distCont} are not valid for conditional settings, which is the case in our study. However, since the tree structure is fixed through time samples, we can define the notion of conditional distance given in Equation \eqref{eq:condDist} as the weighted average of all the individual distances given the covariates.
\begin{equation}
\label{eq:condDist}
 \left[ d_{ij}\vert X\right] := \sum^{K_n}_{k=1} w_{k,ij}d_{k,ij},
\end{equation}
where $w_{k,ij}$'s are the empirical probability matrices of covariate pairs $\left(X_{k,i},X_{k,j}\right)$, such that $\sum_{states}w_{k,ij}=1$, $K_n$ is the total number of observed covariates for each node, i.e.:
\begin{equation}
w_{k,ij} = \Pr(x_{k,i},x_{k,j}).
\end{equation}
For instance,  if the covariates $x_i$ and $x_j$ are binary random variables then their joint has 4 possible states and therefore $w_{k,ij}$ is a $4\times 1$ vector whose entries sum to one.

Individual distances $d_{k,ij}$ for discrete and continuous Gaussian random variables are given in Equations \eqref{eq:discDist} and \eqref{eq:contDist}, respectively. It is worth noting that the additive property of the distance measure will be preserved, due to the fact that the tree is fixed over time and each individual's distance is additive over the tree.

For discrete variables we have:
\begin{equation}
\label{eq:discDist}
 d_{k,ij}:=-\log ( \frac{ \vert {\bf J}(y_i,y_j \vert x_{k,i}, x_{k,j}) \vert}{\sqrt{{\bf M}(y_i \vert x_{k,i}, x_{k,j}) {\bf M}(y_j \vert x_{k,i}, x_{k,j})}}).
\end{equation}
This means that if the covariates are binary then $d_{k,ij}$ will have four states of ``00'', ``01'', ``10'' and ``11'' for the $(x_{k,i}, x_{k,j})$ pair. We then weight each state by the empirical probability state of each covariate pair and average over all covariate pairs and all $k_n$ covariates. This conditional distance measure could be used in, $\estone$ algorithms \cite{choi2011learning} to learn latent graph structure from data.

For continuous variables we have:
\begin{equation}
\label{eq:contDist}
 d_{k,i,j} :=-\log( \frac{ \mathbb{E}(y_i y_j \vert x_{k,i},x_{k,j}) }{\sqrt{\mathbb{E}(y_i^2 \vert x_{k,i},x_{k,j})\mathbb{E}(y_j^2 \vert x_{k,i},x_{k,j})}}).
\end{equation}

\subsection{Parameter Estimation Using EM}
\label{sec:paramEst}
Once we get the latent tree structure using $\clgroup$, we use maximum likelihood to estimate the parameters of the data distribution given in Equation \eqref{eq:expDist} based on the structure. We use Expectation Maximization (EM) to maximize the likelihood function due to the latent node in the structure. EM is an iterative algorithm that iterates between the two so-called E-step and M-step. It maximizes the lower bound of the likelihood function in each iteration based on the parameters estimated in the previous iteration. The lower bound is the expected complete data log likelihood function which is presented in Equation \ref{eq:EM1}.

In order to give a sketch of the EM algorithm formulation, we first present the log likelihood function over the learned latent tree $\Tc_{\dep} = (\Zc_{\dep}, \Ec_{\dep})$:
\begin{align}
\label{eq:likelihood}
	&\ell(\thetabf \vert X, Z) = - \sum_{t=1}^T {\left( A(\thetabf , X^{(t)}) \right)} + \nonumber \\ 
	&\sum_{t=1}^T {\left( \sum_{k \in \Zc_{\dep}}\phi_k( \bfX^{(t)},\thetabf )z^{(t)}_{k} \right)}+ \sum_{t=1}^T {\left(\sum_{kl \in \Ec_{\dep}} \phi_{kl}( \bfX^{(t)},\thetabf )z_{k}^{(t)} z_{l}^{(t)} \right)}
\end{align}
Variable $Z$ is a union of the observed nodes $Y$ and unobserved nodes $H$. Therefore, we cannot maximize the above quantity directly. In order to achieve maximum likelihood we compute the expected complete data log likelihood function given below:
\begin{align}
\label{eq:EM1}
	&\mathbb{E}_{H \vert X, Y}\big( \ell(\thetabf \vert X, Z) \big) =  \sum_{t=1}^T{\left( \sum_{k \in \Zc_{\dep}} \phi_k( X^{(t)})\mathbb{E}_{H \vert X^{(t)}, Y^{(t)}}(z_{k}) \right)} +  \nonumber \\
	&\sum_{t=1}^T{\left( \sum_{kl \in \Ec_{\dep}} \phi_{kl}( X^{(t)},\thetabf )\mathbb{E}_{H \vert X^{(t)}, Y^{(t)}}(z_{k} z_{l}) \right) } - \sum_{t=1}^T{\left( A( X^{(t)},\thetabf) \right)}.
\end{align}
$\mathbb{E}_{H \vert X, Y}$ is computed from the E-step and then the M-step maximizes Equation \eqref{eq:EM1} through gradient descent. 


\section{Experiments and Results}
\label{sec: results}
In order to show the capabilities of our CLTM method, we use 3 different real world datasets.  
The datasets come from two categories, namely \emph{educational data} and \emph{network data}. The educational data is a Massive Online Learning Course (MOOC) dataset~\cite{koedinger2010data} from an online course on psychology offered in Spring 2013. The Network data consists of user interactions and attendance in two social networks. One is an online social media of Twitter and the other one is a network of windsurfers that surf on a beach in Southern California. It should be noted that acquiring large scale dynamic data with long enough time duration that has a reasonable density is a challenging task and most of the available data sets have a very short time duration.

For performance evaluation, we qualitatively observe the estimated tree structures for educational data as the nodes are labeled and can be interpreted. Quantitatively we carry out cross-validation. We learn the model based on the training data and predict nodes/edges evolutions on the test data. A set of scores, which will be defined in the following, are used for performance evaluation. 
We compare our CLTM model with the baseline Chain CRF (CCRF) model in all the experiments in which the chain is over time. We use the same set of covariates for CLTM and CRF for a fair comparison
\begin{figure*}[ht]
        \centering
        \captionsetup{justification=centering}
        \begin{subfigure}[t]{0.3\textwidth}
        \psfrag{Student index}[Bc]{\scriptsize Student index}
        \psfrag{Educational Data}[Bc]{}
        \psfrag{Time index}[c]{\scriptsize Time index}
                \includegraphics[width=\textwidth,height=1.0in]{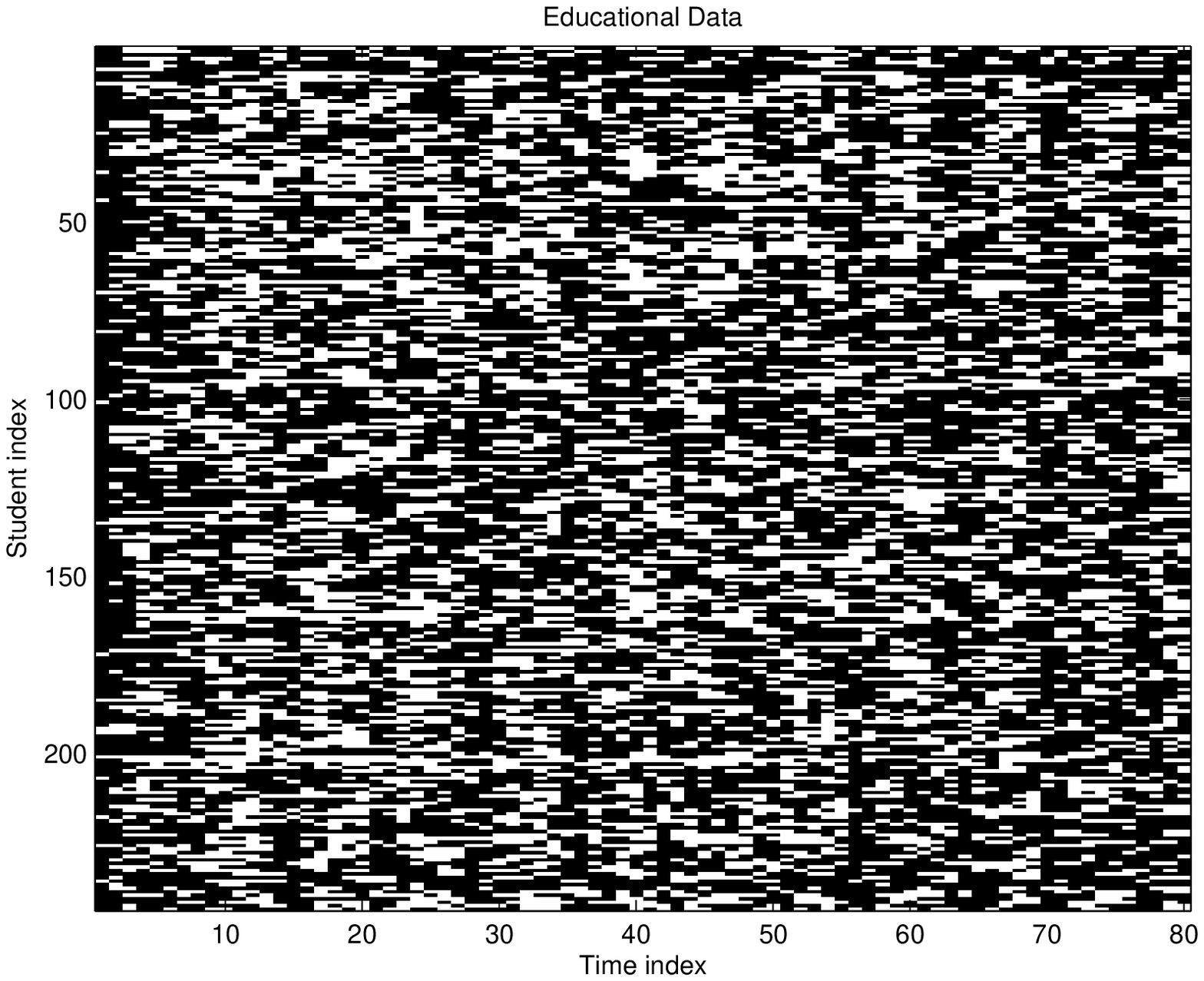}
                \caption{Education data}
                \label{fig:samplesEdu}
        \end{subfigure}
        \begin{subfigure}[t]{0.3\textwidth}
        \psfrag{Vertex index}[Bc]{\scriptsize Vertex index}
        \psfrag{Twitter data vertex samples}[Bc]{}
        \psfrag{Time index}[c]{\scriptsize Time index}
                \includegraphics[width=\textwidth,height=1.0in]{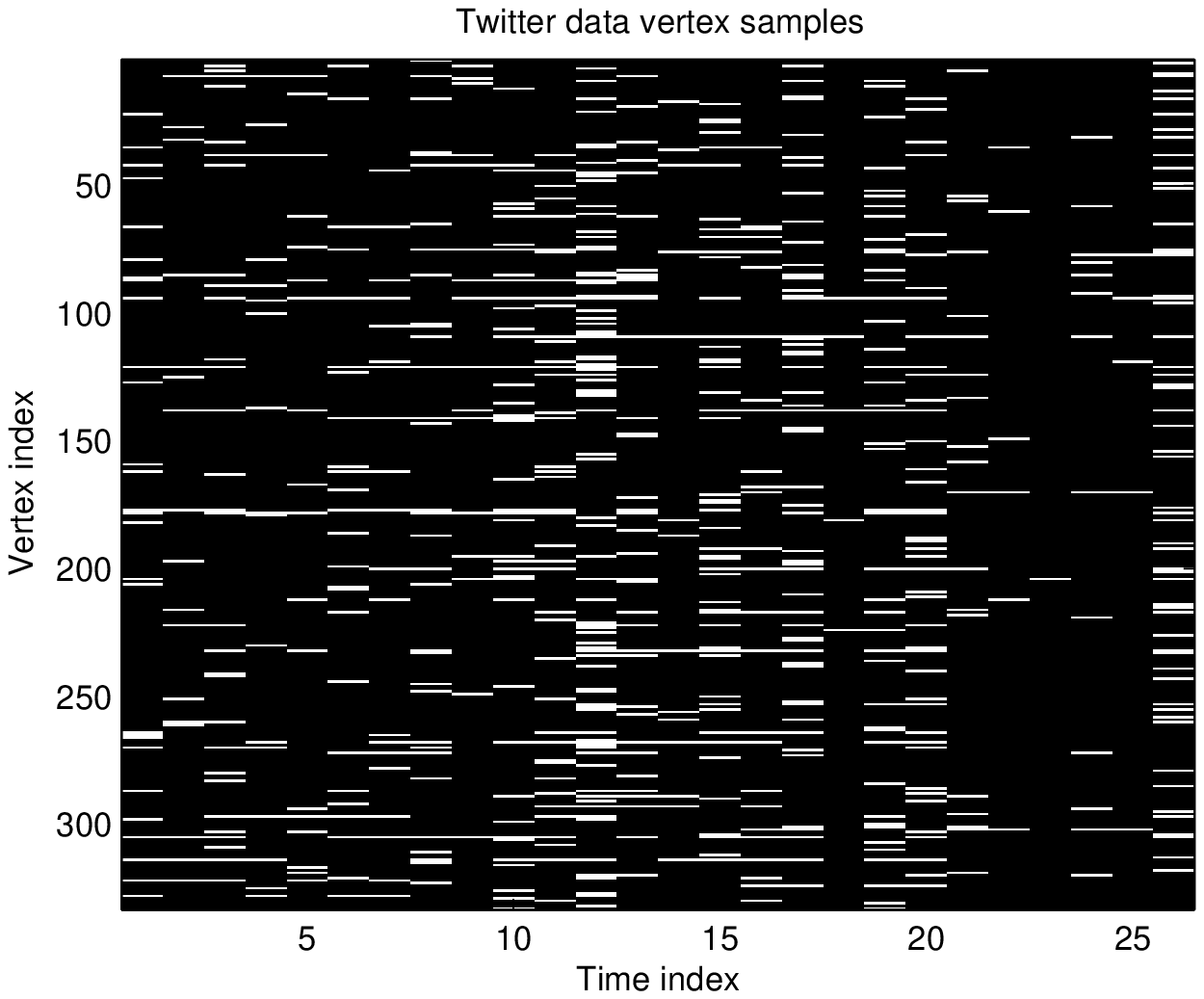}
                \caption{Twitter data}
                \label{fig:samplesTwitter}
        \end{subfigure}
        \begin{subfigure}[t]{0.3\textwidth}
        \psfrag{Vertex index}[Bc]{\scriptsize Vertex index}
        \psfrag{Beach data vertex samples}[Bc]{}
        \psfrag{Time index}[c]{\scriptsize Time index}
                \includegraphics[width=\textwidth,height=1.0in]{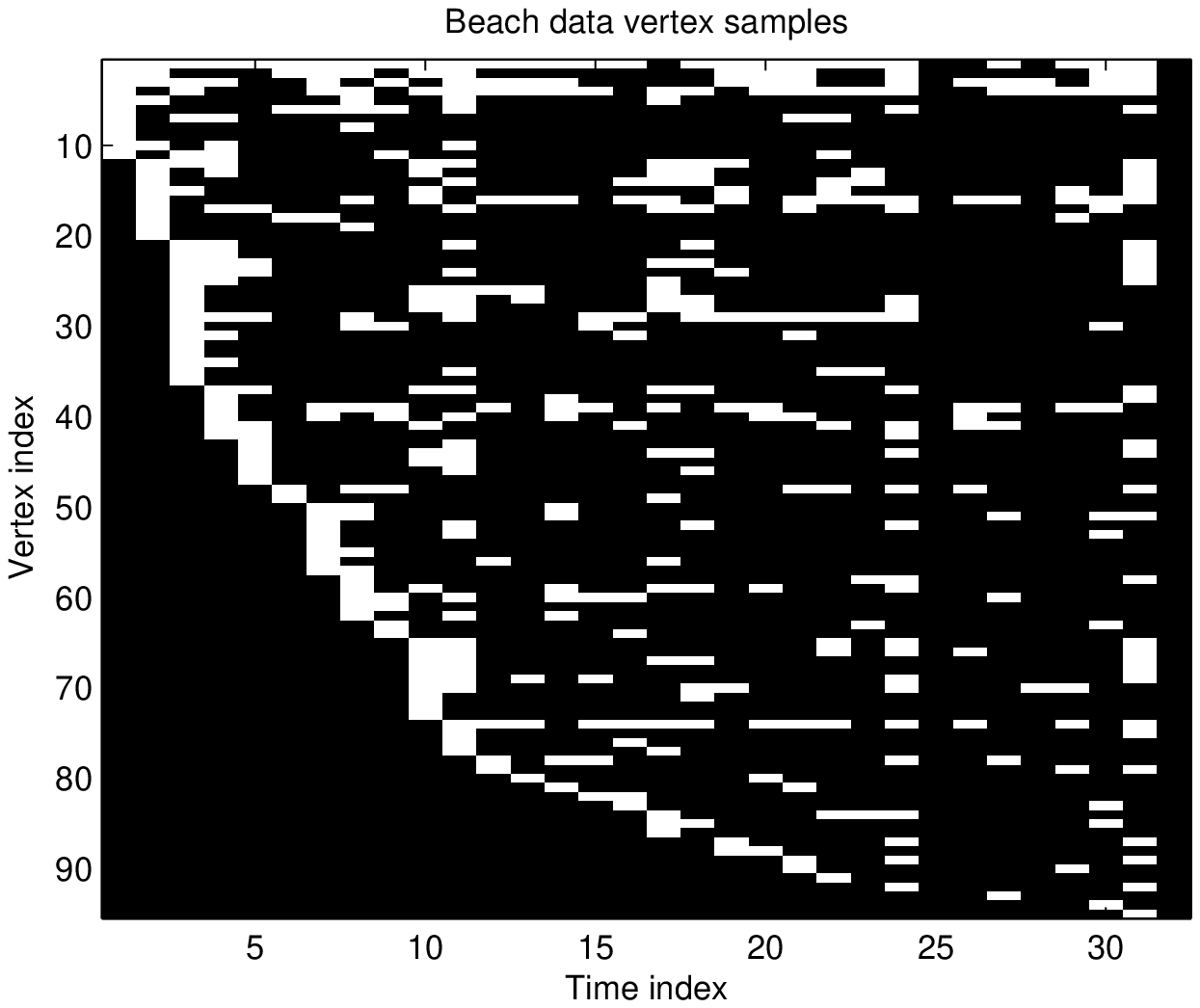}
                \caption{Beach data}
                \label{fig:samplesBeach}
        \end{subfigure}
        \caption{Vertex activities in the datasets. White in education data is correct answer and black is incorrect answer and no activity. In Twitter and beach data, white is active and black is inactive.}
        \label{fig:samples}
\end{figure*}
\paragraph*{Prediction Scores}
\label{sec:perfAnal}
We use the covariates and the estimated parameters of the model to perform one-step-ahead prediction of the data at each time point. 
At each time point, we take the empirical mean of $M$ samples we predict from the estimated model and compare the samples to the ground truth (test data). Repeatedly we carry this out at each time point $t = 1, \dots, T$. 

We define the following measures to assess the performance of the algorithm. 
\begin{compactenum}
\item $\CondP$: the \emph{conditional presence (recall)} which measures the accuracy of predicting a node as active given that the node is indeed active (encoded as 1). 
\item $\CA$: the \emph{conditional absence}, which computes the accuracy of predicting a node's absence (encoded as 0). 
\item $\EAp$: \emph{conditional edge presence (recall)}, where edge indicates a social tie which is oftentimes nodes' interactions. Presence is encoded as 1.  
\item $\EAa$:  the \emph{conditional edge absence}. Absence is encoded as 0.
\end{compactenum}
In the node prediction problem, it is challenging to predict $\CondP$ since the data is highly sparse. Performing well in predicting rare appearances of the data is important. The prediction accuracy is defined as the percentage of nodes predicted correctly out of the $n$ observed nodes. 
In the edge prediction, especially in the case of network data where we want to also track social tie dynamics, we evaluate the prediction accuracy defined as the percentage of edges predicted correctly out of the $e$ possible edges. At the same time, the prediction of absence should not be degraded significantly.

Recall that $y_i^{(t)}$ is observed node $i$ and $w_{ij}^{(t)}$ is the observed social tie between nodes $i$ and $j$ at time point $t=1,\dots,T$. $M$ is the number of predictions drawn from the estimated model. Let $\widehat{y}_{i,k}^{(t)}$ be our $k^{th}$ prediction of node $i$ and  $\widehat{w}_{ij,k}^{(t)}$ be the $k^{th}$ edge prediction between nodes $i$ and $j$ at time point $t$. Let $y_{\text{pred}}^{(t)}$ denote the predicted node set at time point $t$. We define the prediction scores as:
\begin{align}
\CondP(t) &:= \frac{1}{nM}\sum_{k=1}^M\sum_{i=1}^{n}\mathbb{I}(\widehat{y}_{i,k}^{(t)}=1 \vert y_{i}^{(t)}=1),  \label{eq:condPredPresent} \\
\CA(t) &:=\frac{1}{nM}\sum_{k=1}^M\sum_{i=1}^{n}\mathbb{I}(\widehat{y}_{i,k}^{(t)}=0\vert y_{i}^{(t)}=0),\label{eq:condPredAbsent} \\
\EAp(t) &:=\frac{1}{eM}\sum_{k=1}^{M}\sum_{i,j \in y_{\text{pred}}^{(t)}}\mathbb{I}(\widehat{w}_{ij,k}^{(t)}=w_{ij}^{(t)} \vert w_{ij}^{(t)}=1), \label{eqn:edgePresPredAccuracy} \\
\EAa(t) &:=\frac{1}{eM}\sum_{k=1}^{M}\sum_{i,j \in y_{\text{pred}}^{(t)}}\mathbb{I}(\widehat{w}_{ij,k}^{(t)}=w_{ij}^{(t)} \vert w_{ij}^{(t)}=0),\label{eqn:edgeAbsPredAccuracy}
\end{align}
where $n$ is the total number of observed nodes, and $e $ is the total number of possible edges. $\mathbb{I}(.)$ is the indicator function who outputs 1 if its input is true.
 
In addition to the above metrics, we introduce two other measures called \emph{relative difference of average}, $\RDA$, and \emph{relative difference of median}, $\RDM$, that indicate the average and median of the relative improvement of CLTM compared to the baseline model consisting of Chain CRF (CCRF), respectively. 
\begin{align}
\RDA &:=\frac{\sum_{t=1}^{T}\CondP(t)^{CLTM}-\sum_{t=1}^{T}\CondP(t)^{CCRF}}{\sum_{t=1}^{T}\CondP(t)^{CCRF}}, \label{eqn:RDA}\\
\RDM &:=\frac{\text{median}_t(\CondP(t)^{CLTM})-\text{median}_t(\CondP(t)^{CCRF})}{\text{median}_t(\CondP(t)^{CCRF})},\label{eqn:RDM}
\end{align}
The higher the values of RDA and RDM are, the better our performance is compared to Chain CRF. 

\subsection{Educational Data}\label{data:Educational}
The data, gathered from a psychology course in the Stanford Open Learning Library\footnote{available on CMU datashop~\cite{koedinger2010data}}, records students' problem solving outcomes, which are grouped as ``correct'' and ``incorrect''.
The problems that the students answer come in 226 \emph{knowledge components} (KCs).
These knowledge components refer to the different concepts covered in the class throughout the semester.
The multivariate high-dimensional time series data spans 92 days and involves 5,615 students with a total number of 695 problems (2,035 steps). Each problem consists of different steps/stages. There are a total number of 2,493,612 records of students' interactions, which are student's attempts to solve problems, with the server. The course material and problems can be accessed in any order throughout the course. Students' learning behavior and performance are tracked: correct answers are encoded as 1s and the incorrect ones 0s.
The ultimate goal is to track the learning of students and find similar groups of students that behave similarly in terms of learning. First we choose a subset of the students with 244 members.

\begin{figure}[t]
	\centering{
		\psfrag{h1}[Bl]{\scriptsize  ${h_1}$}	
		\psfrag{h2}[Bl]{\scriptsize  ${h_2}$}
		\psfrag{h3}[Bl]{\scriptsize  ${h_3}$}
		\psfrag{x8}[Bc]{\textcolor[rgb]{1,1,1}{\scriptsize $x_{g}$}}
		\psfrag{x7}[Bc]{\scriptsize $x_{h_3}$}
		\psfrag{x5}[Bc]{\scriptsize $x_{1}$}
		\psfrag{x6}[Bc]{\scriptsize $x_{2}$}
		\psfrag{y5}[Bl]{\textcolor[rgb]{1,1,1}{\scriptsize  ${s_1}$}}
		\psfrag{y6}[Bl]{\textcolor[rgb]{1,1,1}{\scriptsize  ${s_2}$}}
		\psfrag{y1}[Bl]{\textcolor[rgb]{1,1,1}{\scriptsize  ${k_1}$}}
		\psfrag{y2}[Bl]{\textcolor[rgb]{1,1,1}{\scriptsize  ${k_2}$}}
		\psfrag{y3}[Bl]{\textcolor[rgb]{1,1,1}{\scriptsize  ${k_3}$}}
		\psfrag{y4}[Bl]{\textcolor[rgb]{1,1,1}{\scriptsize  ${k_4}$}}
		\psfrag{(a)}[c]{\textbf{(a)}}
		\psfrag{(b)}[c]{\textbf{(b)}}
		\psfrag{group 1}[c]{$g_1$}
		\psfrag{group 2}[c]{$g_2$}
	\includegraphics[width = 0.25\textwidth]{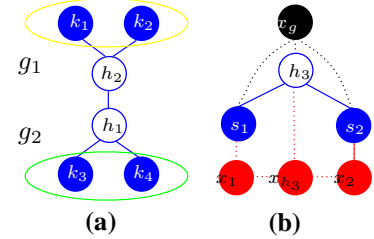}}
	\caption{(a) Presentation of knowledge component latent tree result. In this example 2 groups $g_1$ and $g_2$ are recovered.  (b) Student latent tree structure learning using KC latent tree's grouping as covariates. If student $s_1$ answers problems from knowledge component $k_1$, then $s_1$'s node specific covariate $x_1$ takes value according to $k_1$'s grouping results from (a), which is $g_1$. $x_g$ is the global covariate.}
	\label{fig:KCLT}
	\vspace{-2em}
\end{figure} 

We have two goals for prediction: (1) to learn a latent tree model over the KCs and (2) to predict student learning using KCs as covariates. We first learn a latent tree structure over the KC model that helps us  cluster the concepts~\footnote{The cluster is achieved on the learnt latent tree using standard graph partitioning algorithms~\cite{sen1992graph}}. Then we use these clusters of KC's as relevant covariates to learn a CLTM model for the students' learning behaviors. This process is shown in Figure \ref{fig:KCLT}, where figure (a) shows the latent tree that clusters the KC's, $k_i$, and figure (b) indicates the CLTM structure whose covariates are the clusters of the KC's extracted from figure (a). The ultimate goal is to track the learning of students and find groups of students that behave similarly in terms of learning.

\begin{figure*}[ht]
	\centering{
		\psfrag{concepts humanistic personality}[l]{ \tiny concepts humanistic personality}
		\psfrag{optimism selfefficacy hardiness}[c]{ \tiny optimism self efficacy hardiness}
		\psfrag{relationships support happiness}[l]{ \tiny relationships support happiness}
		\psfrag{relationship money happiness}[l]{ \tiny relationship money happiness}
		\psfrag{drives goals homeostasis intrinsicextrinsic}[c]{ \tiny drives goals homeostasis intrinsic extrinsic}
		\psfrag{affective forecasting}[c]{ \tiny affective forecasting}
		\psfrag{physiology hunger}[l]{ \tiny physiology hunger}
		\psfrag{sexual behavior orientation}[l]{ \tiny sexual behavior orientation}
		\psfrag{eating disorders}[c]{ \tiny eating disorders}
		\psfrag{arousal attraction}[l]{ \tiny arousal attraction}
		\psfrag{critique approaches personality}[c]{ \tiny critique approaches personality}
		\psfrag{inventories tests disorders}[l]{ \tiny inventories tests disorders}
		\psfrag{trait theories personality}[l]{ \tiny trait theories personality}
		\psfrag{anxiety dis gad}[l]{ \tiny anxiety dis gad}
		\psfrag{anxiety dis panic}[l]{ \tiny anxiety dis panic}
		\psfrag{anxiety dis ocd}[l]{ \tiny anxiety dis ocd}
		\psfrag{anxiety dis phobias}[l]{ \tiny anxiety dis phobias}
		\psfrag{anxiety dis ptsd}[l]{ \tiny anxiety dis ptsd}
		\psfrag{dsm mental disorder}[c]{ \tiny dsm mental disorder}
		\psfrag{nuture causes anxiety dis}[l]{ \tiny future causes anxiety dis}
		\psfrag{nature causes anxiety dis}[Bl]{ \tiny nature causes anxiety dis}
		\psfrag{groupthink concepts conditions}[l]{ \tiny groupthink concepts conditions}
		\psfrag{schizophrenia causes}[l]{ \tiny schizophrenia causes}
		\psfrag{schizophrenia symptoms}[l]{ \tiny schizophrenia symptoms}
	\includegraphics[width =0.7\textwidth]{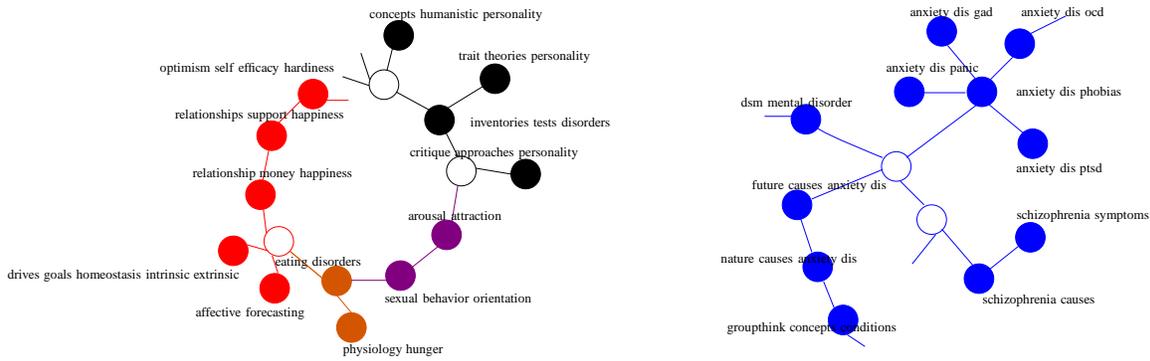}}
	\caption{Subgroups of the estimated knowledge component latent tree. Nodes are colored by the topics of knowledges components. The red nodes talk about relationships and happiness, the black nodes are personality related, the purple nodes are about sexual attractions, and the golden nodes focus on the eating disorders. On the right subgroup, we see a huge cluster of anxiety related concepts, from anxiety disorder all the way to serious schizophrenia symptoms.}
	\label{fig:subs}
	\vspace{-1em}
\end{figure*}

\paragraph{Latent Tree Models for Knowledge Components}The 226 knowledge components have human labels which we use for qualitative interpretation of the learned structure (see Figure \ref{fig:subs}(a) for subset of latent tree learnt, detailed interpretation is in the latter paragraph). The nodes in the learned structure are the KCs and the edges indicate the co-occurrence of the KC pairs in a day for the same student.
Using daily aggregated time points, we consider all students' total numbers of correct and incorrect answers for each of the 226 KCs within a day. The counts are transformed to a ratio by normalizing with the total number of problem solved and are then transformed to approximate Gaussian by taking the square root of the ratio~\cite{brown2008season}. The covariates that we use for learning this structure are some attributes of the data, such as seasonality and previous time points' aggregated outcomes on the KC.

Hierarchical clustering is realized on the knowledge components. The complete learned structure will not fit into the page limits of the paper, but is available online for an interested reader to explore \footnote{Link to the demos will be released in the camera ready version to preserve anonymity of the authors}
We demonstrate two parts of the learned structure in Figure~\ref{fig:subs}. The blank nodes demonstrate the hidden variables learnt, whereas the colored nodes demonstrate the knowledge components. Taking a closer look at Figure \ref{fig:subs} we can see that knowledge components related to  relationships and happiness (red), personality(black), sexual attractions(purple), eating disorders(golden), and anxiety(blue) are clustered together. Thus, we find that our recovered latent tree has considerable face validity with respect to known relationships among topics. Note that none of these labels are input to algorithm, and we require no labels in our unsupervised algorithm.

\paragraph{Predicting Student Learning using learnt groups of Knowledge Components as Covariates} Now that we have the clusters each KC falls into, we use them as covariates along with seasonal information and past observations of the network to track student learning. A subset of students who loyally stayed through the semester (244 members) was selected.
The data is again binned daily and each sample is the ratio of correctly answered problems (aggregating over all KCs) over the total number of problems answered for a student within a day.
We threshold the values to make the data binary. The extracted samples are shown in Figure \ref{fig:samplesEdu} where the horizontal axis indicates time and the vertical axis indicates the students' attempts to answer questions.


\begin{table}[h]
\centering
\caption{Prediction scores for educational data. $\CondP$ and $\CA$ and defined in Equation~(\ref{eq:condPredPresent}-\ref{eq:condPredAbsent}). $\RDA$ and $\RDM$ are defined in Equation~\eqref{eqn:RDA} and~\eqref{eqn:RDM}.}
\label{tab:Res}
\begin{tabular}{|c||c|c|c|c|}
	\hline
	&  $\CondP$ (train) & $\CondP$ (test) & $\CA$ (train) & $\CA$ (test) \\ \hline
	$\RDA$ 	 & 2.6\% 	& 52.96\% & 1.66\% & 2.01\%  \\ \hline
	$\RDM$ 	& 2.68\% 	& 56.37\% & 1.66\% & 1.99\% \\ \hline
\end{tabular}
\end{table}

\begin{figure*}[ht]
        \centering
        \captionsetup{justification=centering}
        \begin{subfigure}[t]{0.8\textwidth}
\psfrag{Parameter value}[Bc]{\scriptsize $\quad \quad \quad$ Covariate Coefficients}
                \includegraphics[width=\textwidth,height = 1.5in]{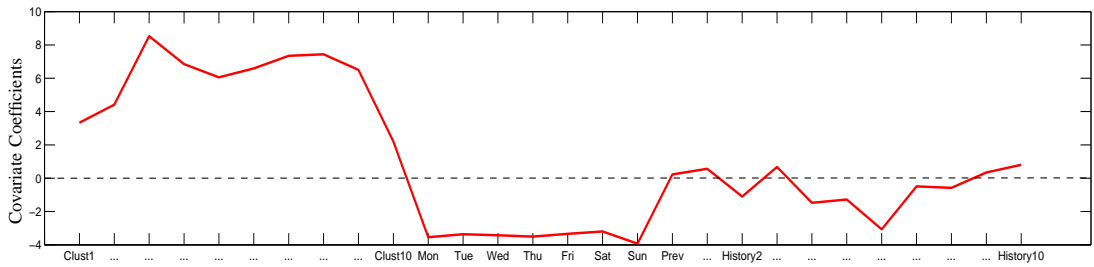}
                \label{fig:ParamsEducation}
        \end{subfigure}
        \vspace{-1em}
        \caption{Covariate coefficients/weights learnt for student learning prediction in education data. }
        \label{fig:ParamsEducationFull}
        \vspace{-1.5em}
\end{figure*}

Educational data covariate coefficients for students is shown in Figure~\ref{fig:ParamsEducationFull}. As illustrated, cluster 3 which is KC groups of ``active early school's contributions to psychology'' and ``physical sci contributions to psychology'' is the most relevant covariate with the highest weight, and smart or hard working students are relevant through those problems. Cluster 9 and 10 which are groups of KCs on ``apply important questions'', ``goals psychology real world'' and ``philosophy contributions to psychology'' are also highly relevant covariates. However, covariates such as KC's on ``brain neuroplasticity'', ``methods studying the brain'' and ``research validity bias '', indicated as cluster 7 in the figure are less relevant in terms of distinguishing student's ability to answer questions in those category correctly.
 It is also interesting to notice that Sunday and Monday happens to be the time that students are most reluctant to work during the week.
Note that the coefficients for seasonality are negative, however this does not imply that the data has negative correlation with seasonality. The reason is that these coefficients are approximately equal, and since at each time point only one of these variables are on, we think of them as a bias term. In other words, seasonality is down-weighting other covariates' effects.

\begin{figure*}[ht]
        \centering
        \captionsetup{justification=centering}
        \begin{subfigure}[t]{0.49\textwidth}
\psfrag{CLTM}[l]{\scalebox{.5}{\hspace{-0.2em}CLTM}}
	\psfrag{LR}[l]{\scalebox{.5}{CCRF}}
	\psfrag{Vertex conditional presence: test}[Bl]{}
	\psfrag{Sccuracy}[c]{\scriptsize Accuracy}
	\psfrag{Time points}[Tc]{\scriptsize Time points}
                \includegraphics[width=\textwidth,height = 1.5in]{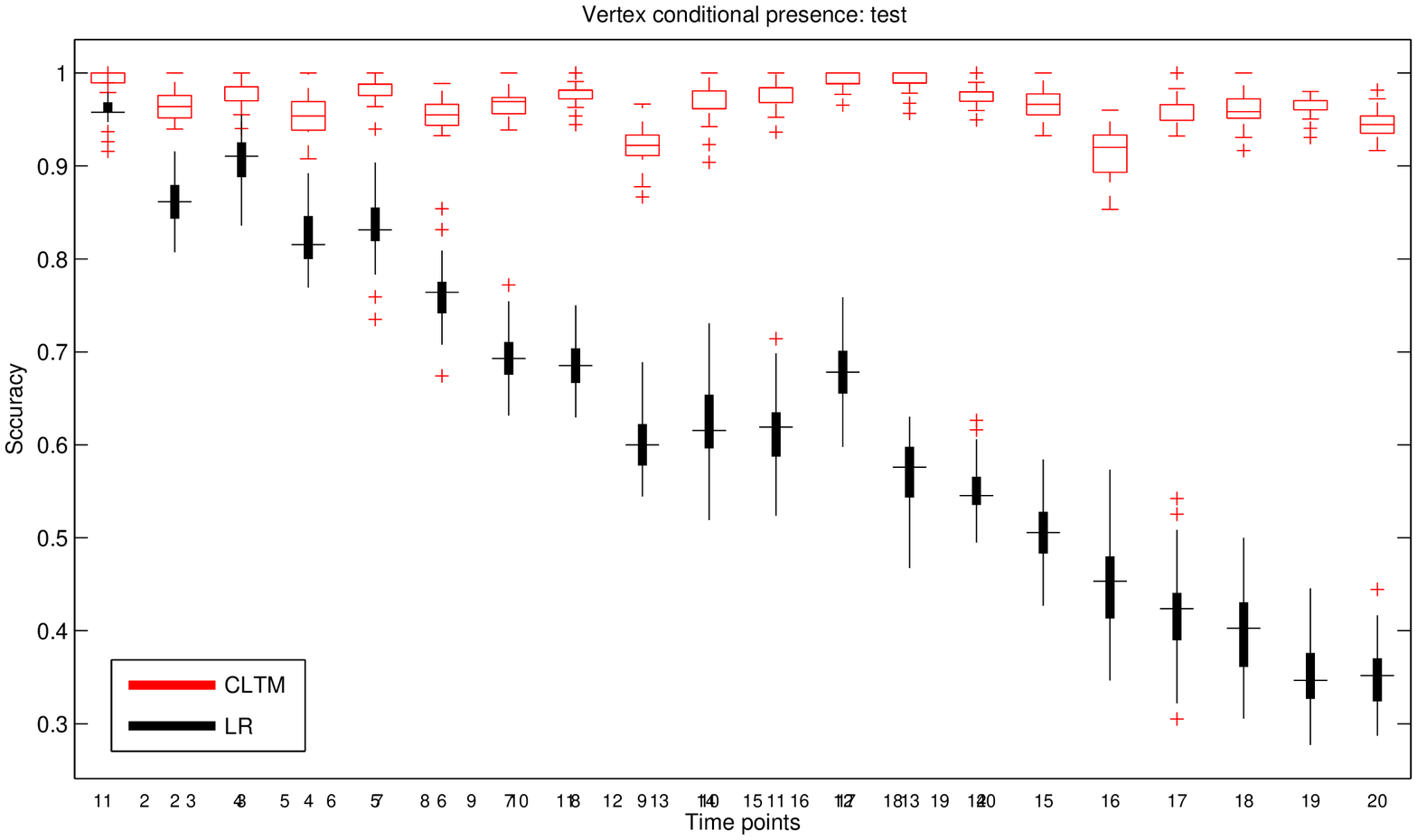}
                \caption{Conditional presence accuracy boxplots for education data on test set}
                \label{fig:EducationCPtestBox}
        \end{subfigure} 
        \begin{subfigure}[t]{0.49\textwidth}
\psfrag{CLTM}[l]{\scalebox{.5}{\hspace{-0.2em}CLTM}}
	\psfrag{LR}[l]{\scalebox{.5}{CCRF}}
	\psfrag{Vertex conditional absence: test}[Tl]{}
	\psfrag{Sccuracy}[c]{\scriptsize Accuracy}
	\psfrag{Time points}[Tc]{\scriptsize Time points}
               \includegraphics[width=\textwidth,height = 1.5in]{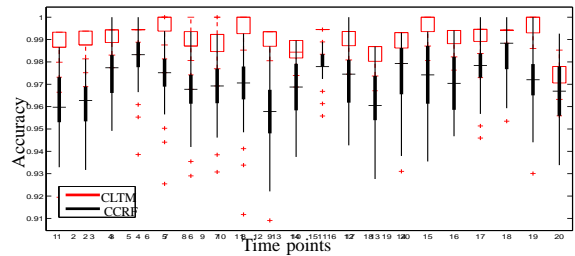}
                \caption{Conditional absence accuracy boxplots for education data on test set}
                \label{fig:EducationCAtest}
        \end{subfigure}
        \vspace{-1em}
        \caption{conditional presence and absence boxplots vs. time for education test data}
        \label{fig:EducationalTest}
        \vspace{-2em}
\end{figure*}
Prediction accuracy curves vs. time, indicating one-step-ahead prediction is presented in Figure~\ref{fig:EducationalTest}. As it is depicted, our algorithm performs significantly better than Chain CRF in the test dataset.

\begin{figure}[ht]
        \centering
        \captionsetup{justification=centering}
        \begin{subfigure}[t]{0.5\textwidth}
        \psfrag{students}[l]{\tiny actual}
        \psfrag{CLTM predictions}[l]{\tiny CLTM predictions}
				\psfrag{LR predictions}[l]{\tiny CCRF predictions}
				\psfrag{Train samples}[c]{\tiny Train samples}
        \psfrag{Test samples}[c]{\tiny \hspace{4em} Test samples}
        \psfrag{student performance}[c]{\tiny \hspace{4em} student performance}
                \includegraphics[width=\textwidth,height=1.2in]{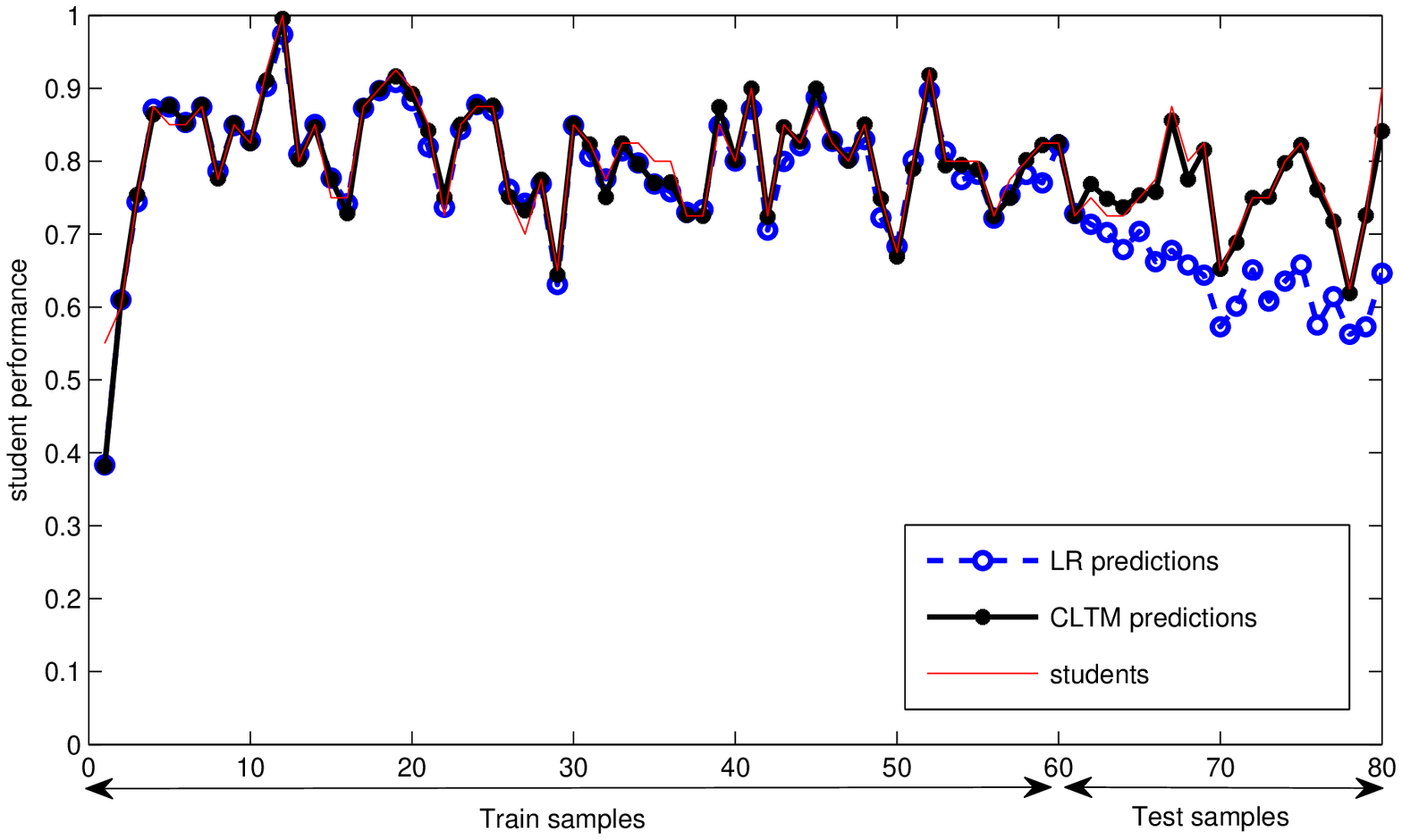}
                \caption{Group of ``strong'' learners}
                \label{fig:smartStd}
        \end{subfigure}
        \begin{subfigure}[t]{0.5\textwidth}
        \psfrag{Train samples}[c]{\tiny Train samples}
        \psfrag{Test samples}[c]{\tiny \hspace{4em} Test samples}
					\psfrag{students}[l]{\tiny actual}
					\psfrag{CLTM predictions}[l]{\tiny CLTM predictions}
					\psfrag{LR predictions}[l]{\tiny CCRF predictions}
					\psfrag{student performance}[c]{\tiny \hspace{4em} student performance}
                \includegraphics[width=\textwidth,height=1.2in]{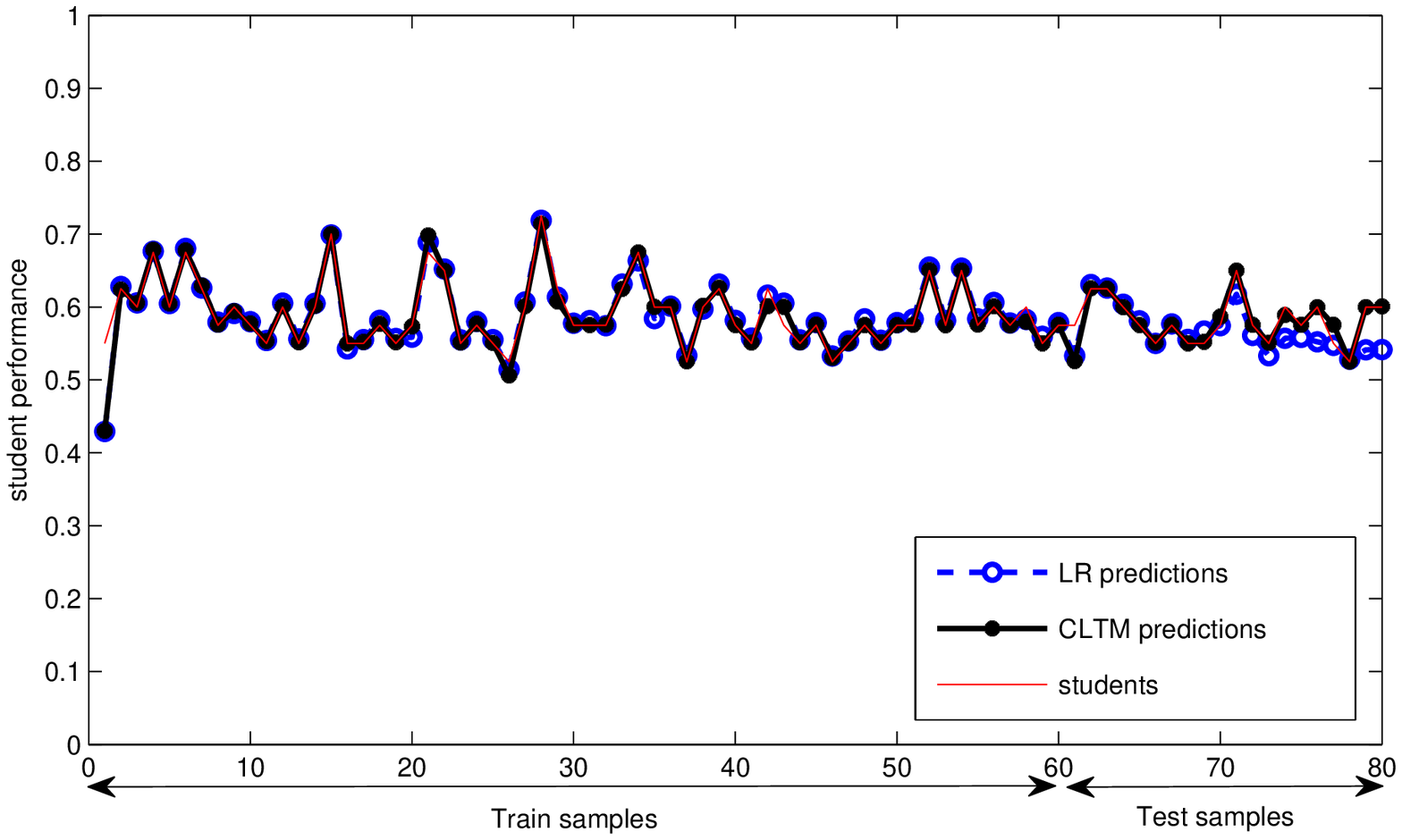}
                \caption{Group of ``weak'' learners}
                \label{fig:dumbStd}
        \end{subfigure}
        \caption{Performance of group of ``strong'' and ``weak'' learners over the semester. First 60 time points are training samples and rest are test. We demonstrate actual performance in answering questions as well as the predictions from the learnt CLTM and CCRF.}
        \label{fig:smartdumb}
        \vspace{-2em}
\end{figure}

Finally, in Figure~\ref{fig:smartdumb}, we demonstrate our ability to automatically find students of similar learning  abilities and track their learning efficiently over time. Among the extracted student groups we choose a group of ``strong'' learners, who accurately answered questions over the time, and ``weak'' ones who were mostly inaccurate in their answers (or inactive and did not answer any questions). We select the group of ``strong'' students who were 1.15 times the average overall performance (over the entire period), and in addition, their neighbors in the conditional latent tree whose information distance is less than the mean distance in the tree. In total, we obtain 21 students for this group. Similarly, for the ``weak'' student group, we consider students who are less than 0.85 times the average performance and also their neighbors in the tree, as described before. In total, we obtain 26 students for this group. We plot the actual performance of these two groups on training and test time periods, as well our predicted performance and the Chain CRF's predictive performance. We see that we closely track the actual performances of these two groups. Also, Chain CRF suffers severely from overfitting as illustrated in this figure. Note that the actual performances of the two groups are significantly different, with the stronger group having much better performance compared to the weaker one.

Thus, our method can automatically find groups of students with similar learning behavior. This can be valuable information for instructors, since they can target these different groups and provide personalized attention of different forms.

\vspace{-1em}
\subsection{Network Data}
This data is gathered from two social networks of beach goers and Twitter users.
In Section~\ref{sec:networkdatades} we present some statistics of the data and give an overview of both datasets.
In Section~\ref{sec:Twitter} we describe the Twitter data and presents its results and compare it to the baseline and in Section \ref{sec:Beach} we talk about the beach dataset and present its results.

\subsubsection{Network Data Description}\label{sec:networkdatades}
The vertex activities for the network datasets are shown in Figures \ref{fig:samplesTwitter} and \ref{fig:samplesBeach}. The horizontal axis indicates time index and the vertical axis indicates vertex index. White represents presence and black represents absence. Both sets of data have similar covariates: we use the previous state of the vertices as well as the number of triads (triplets of nodes which interact with each other) they were engaged in, the previous day as covariates. We also have a covariate that indicates whether the attendee is a regular participant in community activities at the beach, as assessed ethnographically over a period of several months prior to the data collection window. The effect of daily seasonality is also captured by a set of daily fixed effects. We allow for each node to have its own bias indicating that different nodes have different attendance tendency. The positive bias indicates a regular surfer in the beach goers or a regular user among the Twitter users. The negative bias indicates an irregular surfer or user.
Note that in the case of Twitter data, a social tie is defined as direct messaging in Tweets and for the Beach data, a social tie indicates the interaction of surfers while they are at the beach. An overview of specifications of dataset is given in Table~\ref{tab:dataSpec}.
Note that \textsf{AV} and \textsf{AE} denote the average node and edge presence and are indications of the sparsity of the data. 
\begin{table}[h]
\centering
\caption{Data specification and number of used covariates. \textsf{NN}: size of the vertex set, \textsf{AV}: Average Vertex appearance, \textsf{AE}: Average Edge appearance \textsf{NC}: number of node covariates and \textsf{EC}: number of covariates for predicting edges. }
\captionsetup{justification=centering}
\label{tab:dataSpec}
\begin{tabular}{|c||c|c|c|c|c|}
	\hline
	Data & \textsf{NN} & \textsf{AV} & \textsf{AE} & \textsf{NC} & \textsf{EC} \\ \hline
	Education & 244 & 37.17\% & N/A & 29 & N/A\\ \hline
	Twitter &333& 8.49\% & 0.036\% & 9 & 36\\ \hline
	Beach & 94 & 16.66\% & 0.644\% & 11 & 43\\ \hline
\end{tabular}
\end{table}

\begin{table}[h]
\centering
\vspace{-2em}
\caption{Prediction scores for both network data. Conditional presence ($\CondP$), conditional edge presence ($\EAp$), conditional absence ($\CA$) and conditional edge absence ($\EAa$) are defined in Equation~(\ref{eq:condPredPresent}-\ref{eqn:edgeAbsPredAccuracy}). Relative difference of average ($\RDA$) and relative difference of median ($\RDM$) are defined in Equation~\eqref{eqn:RDA} and~\eqref{eqn:RDM}.}
\label{tab:Res}
\begin{tabular}{|c||c|c|c|c|c|}
	\hline
	& Dataset & $\CondP$ & $\EAp$ & $\CA$ & $\EAa$ \\ \hline
	\hline
	$\RDA$ 	& Twitter & 37.66\% 	& 243.73\% & -2.45\% & -0.30\%  \\ \hline
	$\RDM$ 	& Twitter & 42.31\% 	& 565.98\% & -2.85\% & -0.33\% \\ \hline
	\hline
	$\RDA$ 	& Beach  & 10.61\% 	& 60.66\% & -1.98\% & -0.14\% \\ \hline
	$\RDM$	 & Beach &14.10\% 	& 60.81\% & -2.00\% & -0.19\% \\ \hline
\end{tabular}
\end{table}

%

\subsubsection{Twitter Network}
\label{sec:Twitter}
We have collected this dataset by observing the tweeting activity of 333 individuals participating in a discussion on an emergency management topic \#smemchat for a period of 6 months \footnote{We refrain from providing the link to the data to preserve anonymity}. 
The observation period starts from Dec 1st 2013 to Apr 29th 2014
We have a total number of 2313 snapshots of the network which are binned into 26 weekly bins. The vertex set consists of all the nodes that participated in the topic during the observation period and vertex presence is indicated by status updates. Vertex activity peaks on Fridays. Interactions are defined as direct messages among the users, therefore the network is very sparse in terms of user interactions.

A user in the Twitter network is defined as a ``regular'' if he or she appears on the network more frequently than a specified threshold 
A popular user is one whose number of followers is greater than the median of the number of followers of all users. A user is fav if their number of favorites is greater than the median of the number of favorites of all the network attendees. Other covariates remain the same as discussed in the data overview.

\begin{figure*}[ht]
        \centering
        \captionsetup{justification=centering}
        \begin{subfigure}[t]{0.49\textwidth}
        \psfrag{Parameter value}[Bc]{\scriptsize  \hspace{-1.5em} Covariate Coefficients}
	\psfrag{educational data covariate coefficients}[Bl]{}
                \includegraphics[width=\textwidth,height = 1.2in]{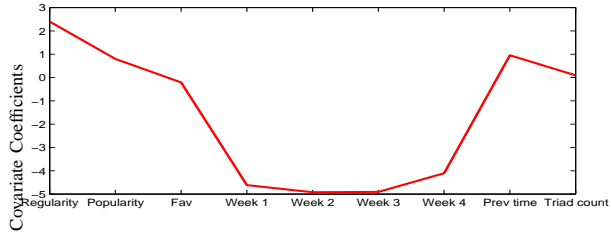}
                \caption{Twitter data covariate coefficients}
                \label{fig:paramsTwitter}
        \end{subfigure}
        \begin{subfigure}[t]{0.49\textwidth}
        \psfrag{Group}[c]{\tiny Group}
        \psfrag{Regularity}[l]{\tiny Reg}
        \psfrag{Monday}[l]{\tiny Mon}
        \psfrag{Tuesday}[l]{\tiny Tue}
        \psfrag{Wednesday}[l]{\tiny Wed}
        \psfrag{Thursday}[l]{\tiny Thur}
        \psfrag{Friday}[l]{\tiny Fri}
        \psfrag{Saturday}[l]{\tiny Sat}
        \psfrag{Sunday}[l]{\tiny Sun}
        \psfrag{Prev time}[l]{\tiny Prev}
        \psfrag{Triads}[l]{\tiny Triads}
        \psfrag{Time points}[Bc]{\footnotesize Time Points}
        \psfrag{Parameter Value}[c]{\scriptsize  \hspace{-0.5em} Covariate Coefficients}
                \includegraphics[width=\textwidth,height = 1.2in]{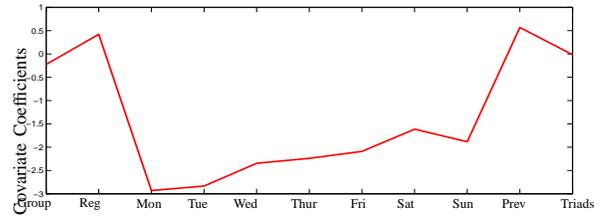}
                \caption{Beach data covariate coefficients}
                \label{fig:paramsBeach}
        \end{subfigure}
        \vspace{-1em}
        \caption{Node covariate weights/coefficients learnt for predicting node participation in Twitter and beach data. } 
        \label{fig:ParamsTwitterT}
        \vspace{-1em}
\end{figure*}

The weights learned for the covariates described is shown in Figure~\ref{fig:paramsTwitter}. As illustrated, regularity of the user and its past time activity are the most relevant covariates with the highest weights. Seasonality covariates are the week number in each month. As expected different weeks within each month behave approximately similar. Also, as it was the case with Educational data, seasonality covariates are down-weighting the other covariates' effects rather than having negative correlation with data.

\begin{figure*}[ht]
        \centering
        \captionsetup{justification=centering}
        \begin{subfigure}[t]{0.49\textwidth}
        \psfrag{edge covariate coefficients}[Bl]{}
        \psfrag{Parameter value}[Bc]{\scriptsize  \hspace{-1.5em} Covariate Coefficients}
                \includegraphics[width=\textwidth,height = 1.2in]{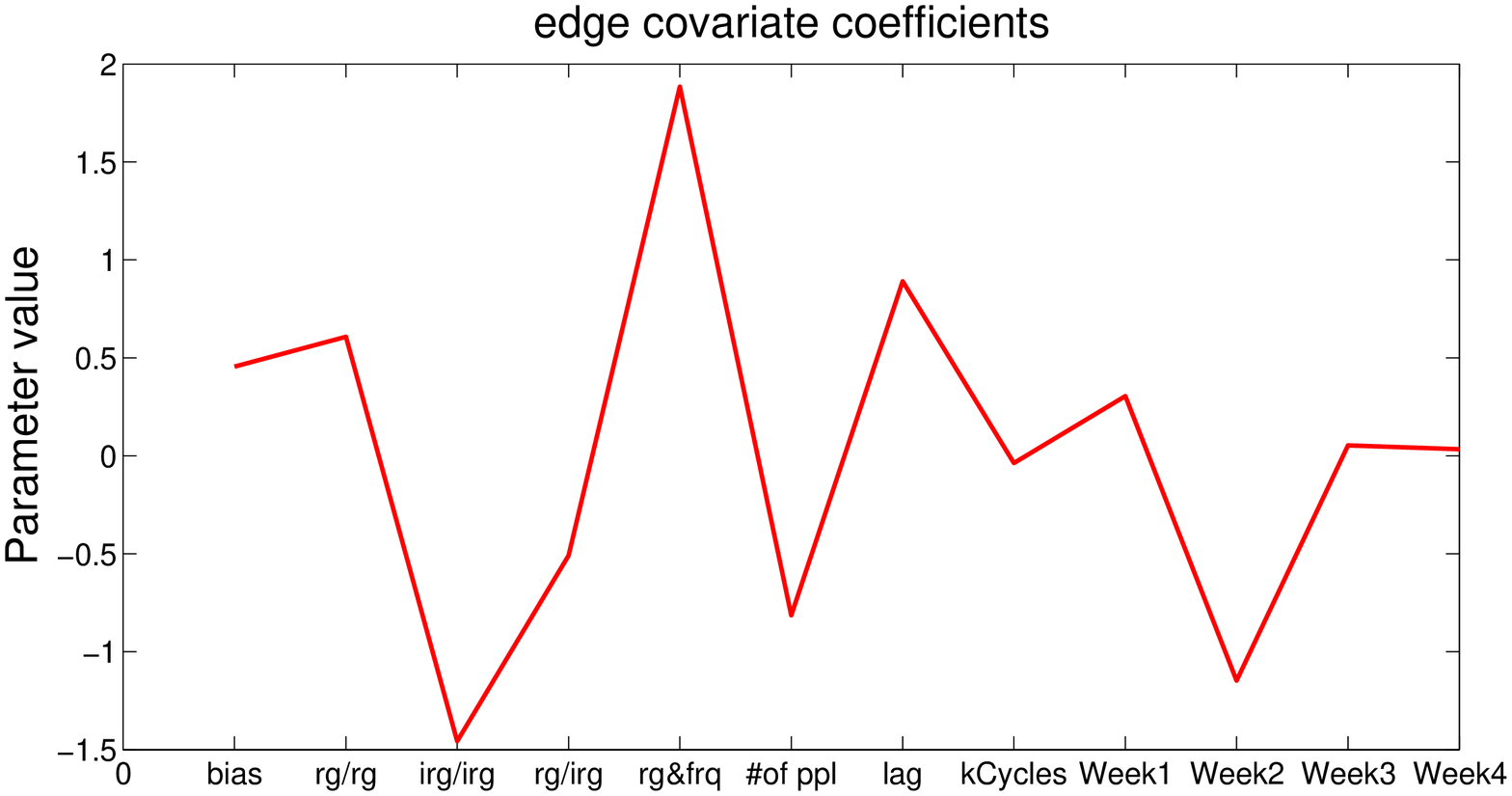}
                \caption{Coefficients of covariates for edge prediction}
                \label{fig:EparamsTwitter}
        \end{subfigure}
        \begin{subfigure}[t]{0.49\textwidth}
        \psfrag{hidden node state regressors}[Bl]{}
        \psfrag{Parameter value}[Bc]{\scriptsize \hspace{-2em}  Node Specific Bias}
        \psfrag{Hidden node index}[Bc]{\scriptsize{\hspace{0.7em}Hidden node index}}
                \includegraphics[width=\textwidth,height = 1.2in]{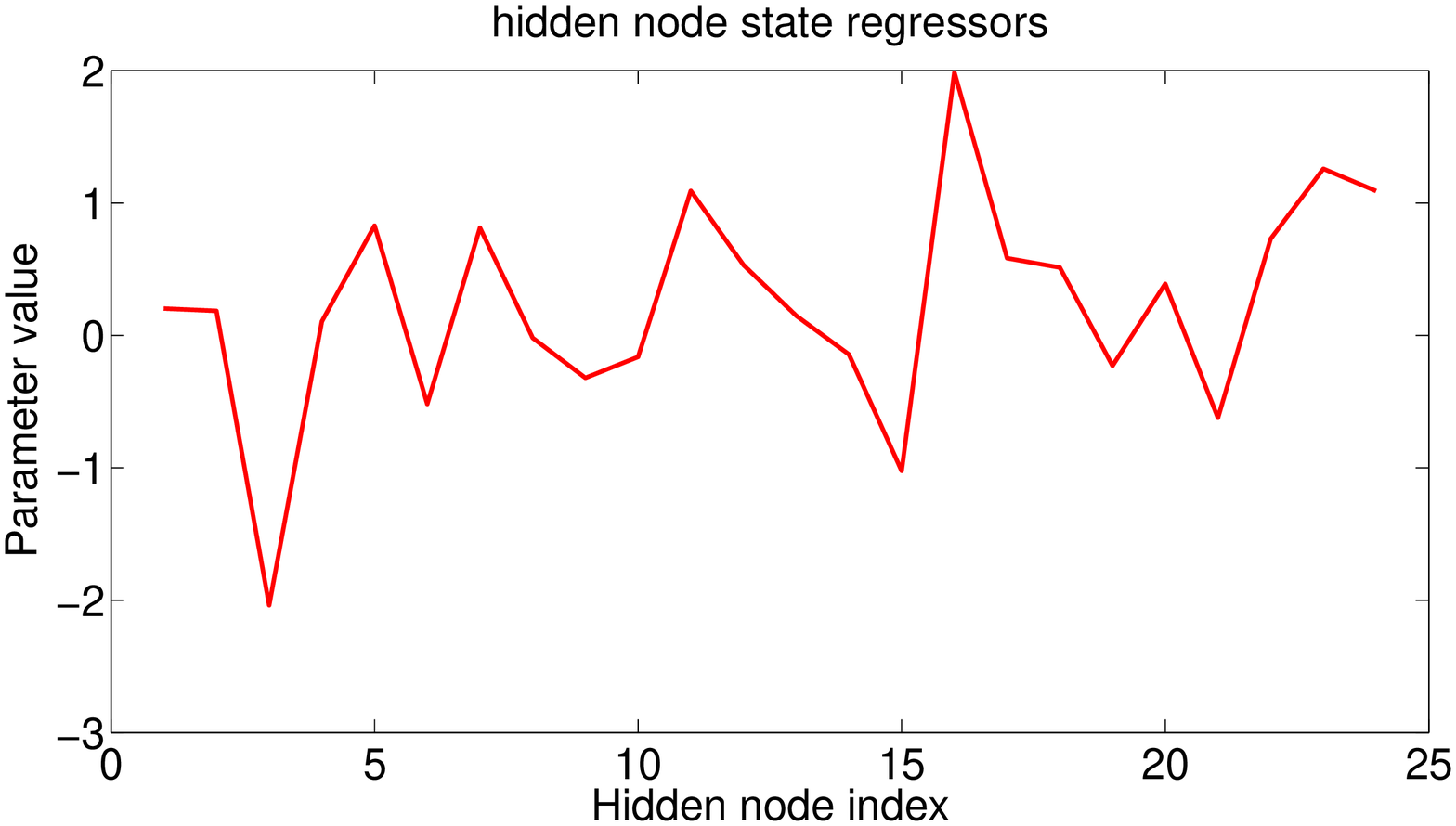}
                \caption{Coefficients of covariates based on inferred hidden states of the vertex CLTM, used for edge prediction} 
                \label{fig:EParamsTwitterHidden}
        \end{subfigure}
        \vspace{-1em}
        \caption{Learnt coefficients for various covariates for edge prediction in Twitter data}
        \vspace{-1em}
        \label{fig:ParamsTwitterTie}
\end{figure*}

As mentioned in Section \ref{sec:EdgePred}, we are also interested in tracking social interaction dynamics conditioned on the predicted node state of the network using the inferred state of the hidden variables in the CLTM model. The idea is that an edge cannot form unless both nodes that form it are present. Following the same reasoning for prediction, we limit edge prediction to the predicted node set; which drastically reduces the sparsity of the edge set.
The covariates used for edge prediction are seasonality effects, previous state of the network, number of present nodes in the previous day, number of K-cycle structures~\cite{may1976simple}, the presence of the edge  in the previous day, whether the  edge is between  regular-regular, regular-irregular or irregular-irregular network users. The coefficients learned for these covariates are shown in Figure \ref{fig:EparamsTwitter}.
We find  that regular/regular interactions (interactions between regular users) have a high weight, whereas the weight of an irregular/irregular interaction is very low indicating that regular nodes are more likely to talk to one another compared to irregular nodes. Also as expected, regular/irregular interactions are somewhere in between the two.
Another interesting point is that if one of the nodes that form a specific edge are a frequent and regular attendee of the network, they will have a very important role in prediction. Previous time state of the network is also another thing that highly affects prediction.


Figure~\ref{fig:EParamsTwitterHidden} shows the learnt weights for the covariates which consist of the inferred hidden variables from the vertex conditional latent tree model. In other words, after the CLTM is learnt over the vertices, the configuration of hidden nodes is inferred through belief propagation. These inferred values are then incorporated as covariates into the edge model as follows: for predicting each edge, we incorporate the configuration of hidden variables which are parents of the endpoints of the corresponding vertices that form the edge. In Figure~\ref{fig:EParamsTwitterHidden}, we see that different hidden nodes affect the presence of edges to different extents, thus indicating that different groups have varying tendencies for forming edges.

\begin{figure*}[!ht]
        \centering
        \captionsetup{justification=centering}
        \begin{subfigure}[t]{0.49\textwidth}
        \psfrag{CLRF CP}[Bl]{\scalebox{.5}{\hspace{-0.2em}CLTM $\CondP$}}
        \psfrag{DNR CP}[Bl]{\scalebox{.5}{\hspace{-0.4em}CCRF $\CondP$}}
        \psfrag{CLRF CA}[Bl]{\scalebox{.5}{\hspace{-0.2em}CLTM $\CA$}}
        \psfrag{DNR CA}[Bl]{\scalebox{.5}{\hspace{-0.4em}CCRF $\CA$}}
        \psfrag{Accuracy}[Bc]{\scalebox{.9}{\hspace{2em}Accuracy}}
        \psfrag{Time points}[Bc]{\scalebox{.9}{\hspace{-0.5em}Time Points}}
        \psfrag{Vertex conditional presence and absence}[l]{}
                \includegraphics[width=\textwidth,height = 1.5in]{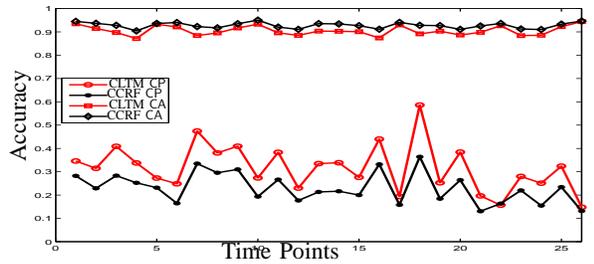}
                \caption{Vertex prediction accuracy}
                \label{fig:TwitterVertex}
        \end{subfigure} 
        \begin{subfigure}[t]{0.49\textwidth}
        \psfrag{CLRF EP}[Bl]{\scalebox{.5}{\hspace{-0.2em}CLTM $\EAp$}}
        \psfrag{DNR EP}[Bl]{\scalebox{.5}{\hspace{-0.4em}CCRF $\EAp$}}
        \psfrag{CLRF EA}[Bl]{\scalebox{.5}{\hspace{-0.2em}CLTM $\EAa$}}
        \psfrag{DNR EA}[Bl]{\scalebox{.5}{\hspace{-0.4em}CCRF $\EAa$}}
        \psfrag{Accuracy}[Bc]{\scalebox{.9}{\hspace{2em}Accuracy}}
        \psfrag{Time points}[Bc]{\scalebox{.9}{\hspace{-0.5em} Time Points}}
        \psfrag{Edge conditional presence and absence}[l]{}
                \includegraphics[width=\textwidth,height = 1.5in]{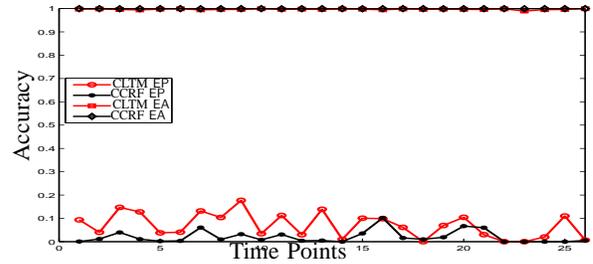}
                \caption{edge prediction accuracy}
                \label{fig:TwitterEdge}
        \end{subfigure}
         \vspace{-1em}
        \caption{Vertex and edge prediction accuracy for the Twitter dataset}
        \label{fig:TwitterVertexEdge}
        \vspace{-2em}
\end{figure*}

Prediction accuracy curves vs. time, indicating one-step-ahead prediction for both node prediction and edge prediction are also presented in Figure~\ref{fig:TwitterVertexEdge}, respectively. As it is depicted in the Figure~ \ref{fig:TwitterVertex}, improved vertex prediction accuracy boosts edge prediction performance, since edge prediction is conditioned on the predicted node set. Incorporating the inferred state of the hidden variables of the CLTM model is another important factor that increases prediction accuracy compared to the baseline Chain CRF without the inferred states.
As illustrated in Figure~\ref{fig:TwitterEdge}, CLTM improves $\CondP$ while maintaining a good $\CA$, resulting in a 243.73\% improvement in average $\EAp$.

\subsubsection{Beach Network}
\label{sec:Beach}
This data contains a dynamically evolving network of interpersonal communications among individuals congregating on a beach in Southern California observed over a one-month period \cite{freeman1988human}. The vertex set in this network is the windsurfers appearance on the beach in this 31 day period and the edge set is composed of their interpersonal communications recorded in the data set.

The network was tracked two times a day, for 31 days from Aug 28, 1986 to Sept 27, 1987 by Freeman et. al. There is a total number of 94 windsurfers who are divided into two groups of \emph{regulars} (with 54 members) and \emph{irregulars} (with 41 members). The groups of \emph{regulars} is further categorized into two groups of ``Group 1'' with 22 members, ``Group 2'' with 21 members, leaving 11 individuals in this category as ungrouped. Vertex appearance on the beach ranges from 3 to 37 in the 31 day tracking period. The number of communication ties per day ranged from 0 to 96 in this dataset.

The covariates used by the vertex model are the regularity effect, group terms and all other covariates described in the network data description. The covariate weights learned by the algorithm is illustrated in Figure \ref{fig:paramsBeach}. The highest weight is given to the previous vertex state and regularity. The same discussion about the effect of seasonality to the data holds for the Beach data as well. But here we can see that Saturday has a slightly higher weight than the rest of seasonality covariates, and this indicates that Saturday down-weights the other parameters less than the other days, which in turn means that it is more likely that people come to the beach if it is a Saturday.

\section{Conclusion}
\label{sec: conclusion}
In this paper we propose a framework for modeling and tracking high-dimensional temporal data via conditional random fields. An approximation of latent tree structure in the conditional random field allows for efficient inference. This is a step forward towards understanding the high-dimensional time series with covariate effects. The success application of our proposed method to educational data and network data reveals potential in applying the method to a wider class of problems. 

\vspace{-1.5em}
\section*{Acknowledgements}
F. Arabshahi and F. Huang are supported by NSF BIGDATA grant no. IIS-1251267. F. Arabshahi was also partially supported by the UCI Data Science Initiative summer fellowship. A. Anandkumar is supported in part by Microsoft Faculty Fellowship,
NSF Career award CCF-1254106,  ONR Award N00014-14-1-0665, ARO YIP Award W911NF-13-1-0084 and AFOSR YIP  FA9550-15-1-0221. C. T. Butts and S. M. Fitzhugh are supported in part by NSF grants OIA-1028394, CMMI-1031853 and IIS-1251267 and Army Research Office award ARO-W911NF-14-1-0552. 

%
\IEEEpeerreviewmaketitle

\end{document}